# Reconnection in the Centre of the Near-Earth Plasma Sheet: Control of Onset by Plasma Drifts and Magnetic Flux convection close to the earth

G. Atkinson
Department of Physics and Astronomy, University of British Columbia
Vancouver, B.C. V6T 1Z1, Canada
atkinson@chem.ubc.ca

**Abstract**.  Convection inside the magnetosphere can be regarded as the transfer of magnetic flux and plasma from the magnetotail into the nightside and then convection and drift from there to the dayside.  As in many fluid flow situations, the downstream flow (the convection and drift to the dayside) is an important boundary condition for the upstream flow (reconnection and inflow from the tail).  Tail-like flux tubes in the near-earth plasma sheet block inflow from the tail between reconnection events.  However, the westward drift of the energetic ions (curvature and gradient drift and **ExB** drift) removes plasma energy from near midnight causing deflation and earthward flow in the equatorial plane.  (The process can be described as the westward drift of the eastern of the plasma sheet / partial ring current.)  The earthward flow removes magnetic flux and plasma from the center of the thick plasma sheet and creates a thin current sheet, which, in turn, favors the onset of reconnection.  The X line forms just tailward of the earthward-flow region.  Initially, reconnection occurs by the earthward propagation of the X line and the growth of a magnetic island on its tailward side, since there is no downtail flow.  Subsequently, the system evolves towards lobe merging, and the X line propagates tailward consistent with the standard picture of tail reconnection.  The east-west width of the local-time slot in which inflow and reconnection occur is determined by the scale-length of the decrease in plasma energy at the eastern end of the drifting ions  It is expected that the reconnection slot follows the westward drift of the eastern end of the partial ring current.  The model appears to be consistent with bursty bulk flows, the westward traveling surge and many other observed properties of substorms.

## 1. Introduction

The interaction of the solar wind with the magnetosphere results in the transport of magnetic flux into the magnetotail (Dungey, 1961; Axford and Hines, 1961).  To avoid a continued buildup in the tail, there is a return convection of magnetic flux from the magnetotail into the nightside dipole-like region and from there to the dayside.  The magnetic flux transfer rate through any line is given by $V = \int E \cdot dl$ where $dl$ is a line element and $E$ is the electric field vector in a reference frame fixed with respect to the line.  We will be interested in flux transfer rates both at ionospheric heights and in the equatorial plane of the magnetosphere, so that the lines of interest lie in one of these two surfaces.

The magnetic flux transfer is complicated by the presence of energetic plasma.  There is too much energy on nightside tail-like plasma sheet flux tubes for a simple earthward convection from the tail into the dipole-like region (and subsequently to the dayside).  This problem has become known as the "pressure problem" or "pressure catastrophe" (Erickson and Wolf, 1980; Schindler and Birn, 1982).  Magnetic flux cannot be "dipolarized" by adiabatic earthward convection of tail-like magnetic flux tubes; in fact, earthward convection results in the field becoming more tail-like.

It is generally believed that substorms are the magnetosphere's solution to the pressure problem.  A commonly-accepted model of substorms (e.g. Hones, 1984; Baker et al. 1996) is the following sequence: 1) The appearance of a strong southward component to the interplanetary magnetic field increases the dayside reconnection rate and transport of magnetic flux into the magnetotail. 2) The pressure problem limits transport of magnetic flux from the tail into the nightside magnetosphere and hence there is a buildup of magnetic flux in the tail. 3) The increasing pressure of the magnetic flux in the lobes causes the plasma sheet to become thinner and the field lines more tail-like. 4.) Eventually the plasma sheet becomes thin enough that MHD conditions break down and reconnection is initiated at a near-earth neutral line. 5) The energy problem is solved by the ejection of plasma and energy in the downtail direction leaving behind more-dipolar magnetic flux. There have been numerous studies of the onset of reconnection due to the plasma sheet becoming thin and of the resulting behavior of the magnetotail (e.g. Zhu and Winglee, 1996).

The above model has been successful in explaining many features of substorms, but there are some which are more difficult to include in the model (they are not necessarily inconsistent, but they do not result directly from the theory). Among these are: convection without substorm expansions (Nishida, 1971; Kamide and Kokubun, 1996), the detailed behavior of auroral arcs, the triggering of substorms by northward turnings of the solar-wind magnetic field (e.g. McPherron et al., 1986; Rostoker et al., 1983), the equal occurrence frequency of earthward and tailward moving X lines (Ueno et al., 1999), the existence of two centers of activity associated with substorm onset (at ~10Re and ~25Re downtail) (e.g. Lui et al., 1988; Shiokawa et al., 1992), and the role of bursty bulk flows in substorms and in the large-scale magnetospheric convection (e.g. Baumjohann et al., 1990; Schodel et al., 2001; Angelopoulous et al., 1992).

In this paper, a modification of step 4 in the above standard substorm sequence is proposed. It is argued that the outflow of plasma and magnetic flux from the nightside to the dayside in the near-earth plasma sheet is a boundary condition that is essential to the physics of inflow from the tail and the onset of reconnection at a near-earth X line. Thus step 4 is modified by the argument that the onset of reconnection depends strongly on the flow out of the nightside in the near-earth plasma sheet.

Since near-earth convection and drift cannot be neglected in considering reconnection and flow into the nightside, a model of a convecting magnetosphere is essential to the study. A simplified three-region model is used. Each region is composed of magnetic flux with different plasma properties. The author believes that the three-region model contains the minimum physics required to reproduce most of the observed features of large-scale magnetospheric convection.

The paper is divided into 8 sections. In section 2, it is argued that convection of plasma and magnetic flux out of the nightside is a boundary condition for near-earth reconnection and inflow from the tail that cannot be ignored. In section 3, the three-region magnetosphere is discussed. In section 4, the boundary between the two highest-latitude regions is discussed in detail, including the requirement for reconnection at a "critical" X line located at this boundary. In section 5, it is argued that earthward flow (caused by the removal of plasma energy earthward of the boundary) will lead to reconnection at a critical X line. In section 6, the control of the onset of reconnection by earthward flow is illustrated with a "thought experiment". In section 7, the evolution of reconnection after its onset, and the production of bursty bulk flows are discussed. Section 8 is a summary of the paper.

## 2. Control of nightside reconnection by convection to the dayside

Convection inside the magnetosphere and tail can be compared to the flow of fluid through a pipe containing two "blockages". The pressure drop across each is determined by the requirement that the flow rates through the two blockages be equal, for an incompressible fluid; or that the time average of the flow rates be equal for a compressible fluid. Time variations of the blockage at the downstream obstruction cause variations of the flow through the upstream blockage. The downstream flow is a boundary condition for flow through the upstream blockage which cannot be neglected unless the downstream blockage is much weaker, and the pressure drop across it is negligible.

Convection from the nightside tail to the dayside magnetosphere involves two "blockages". The upstream blockage is provided by the nightside plasma sheet and its current system which limits the flow of magnetic flux and plasma from the tail into the nightside dipole-like region. The downstream blockage is provided by the ionospheric line-tying currents which limit the convection out of the nightside toward the dayside. Since storage of the fluid (plasma plus magnetic flux) can occur in the nightside near-earth plasma sheet between the two blockages, the system is similar to compressible flow through a pipe. The time average of the two magnetic flux transfer rates must be equal. It can be concluded that the convection rate from the nightside to the dayside is a boundary condition that cannot be neglected in considering reconnection and inflow in the near-earth tail.

It is clear that a model of the global magnetospheric convection is essential for the study of nightside near-earth reconnection and inflow.

## 3. The three-region magnetosphere: the three regions, the two boundaries and convection

A simplified model of the magnetosphere must contain at least three regions of magnetic flux (e.g. Atkinson, 1994; 2002). Each region has different plasma properties and the two boundaries between the regions can be treated as discontinuities. Figure 1 illustrates the three regions. The plasma energy determines that the stable configuration is with region A at the highest latitudes and region C at the lowest.

Region A is the highest-latitude region. It consists of both tail-lobe and tail-like plasma-sheet flux tubes. We define it as magnetic flux which is too tail-like to convect to the dayside without removal of energy by reconnection or other process. Azimuthal convection of region A flux around the earth to the dayside is prevented by the flanks of the magnetosphere, essentially by the solar wind, and earthward convection is prevented by the plasma pressure gradient.

Region B consists of magnetic flux with feet at auroral latitudes. It includes the near-earth plasma sheet and ring current. We define region B as magnetic flux which contains enough plasma energy to have a major effect on convection, but which can convect to the dayside without the removal of plasma and energy. (Note that this definition is not the same as in earlier works by the author.) Region B contains the azimuthal convection from the nightside to the dayside that is observed at auroral latitudes.

Region C is the lowest-latitude region and consists of dipole-like flux tubes without significant plasma pressure. We define region C as magnetic flux that can convect azimuthally around the earth, and that does not contain significant plasma pressure.

The lower-latitude boundary (B-C) lies between regions B and C. It is the earthward edge of the plasma-sheet and ring-current plasma, and of the azimuthal convection around the earth toward the dayside. It is the location of the region 2 Birkeland currents.

The higher-latitude boundary (A-B) is between regions A and B. It will be discussed in some detail in the next section.

Consider the large-scale convection. In a static non-convecting magnetosphere, all currents close in the magnetosphere. Any distortion from the static configuration results in convection and closure of some of the current to the ionosphere. Thus it is the distortion from the static configuration which drives convection. Distortions are produced by dayside and nightside reconnection, as well as other processes such as solar wind pressure changes.

An example of this is the simplest-possible three-region magnetosphere, with axial symmetry for the static non-convecting situation. The magnetospheric currents follow circular paths around the axis of symmetry. Perturbations from axial symmetry lead to closure of some of the current to the ionosphere, and hence convection. Perturbations can be produced by flux transfer between the regions, and appear as perturbations of the boundaries A-B and B-C from the circular shape expected from axial symmetry.

If we apply the above considerations to substorms, a growth phase begins when a large increase in dayside reconnection starts to distort the magnetosphere from the quiescent configuration by transferring dayside flux from region B to region A. The degree of distortion increases with time and so does the azimuthal convection to the dayside in region B. At expansion onset, a nightside reconnection event transfers flux from region A to region B on the nightside. The result is an additional distortion of the magnetosphere, the effects of which are added to the previous distortion. In general, the distortion of the magnetosphere at a given time (and hence the global convection) is the sum of effects due to dayside and nightside reconnection in the recent history of the magnetosphere. (Solar wind pressure variations have not been considered in this discussion).

In the next section of the paper, we look at flux transfer events from region A into region B on the nightside.

## 4. The boundary A-B and critical X lines

The higher latitude boundary (A-B) is of major importance to the convection because of the above definitions of regions A and B. We defined region B as flux that can convect to the dayside and region A as flux that cannot. It follows that convection from the magnetotail to the dayside can occur only if there is reconnection at an X line located near A-B for at least some of the time. This location for an X line is illustrated in figure 1b. X lines can also form further downtail producing mid-tail plasmoids (e.g. Nishida et al, 1986), but these do not produce flux that can convect to the dayside. We shall refer to X lines that produce region B flux as "critical" X lines.

Another way of looking at the necessity for critical X lines in the magnetosphere is that if X lines were forming only at locations well downtail of A-B at some initial time, reconnection would shorten region A field lines but not produce flux capable of convecting to the dayside. The result would be a continued buildup of region A flux until the site of X-line formation moved to A-B. That is, the magnetosphere would change topology until the average flux transfer rates through A-B and to the dayside were equal to the average dayside reconnection rate.

The location of A-B and of the critical X line is of great interest. Can it be ~25 Re downtail from the earth consistent with the reported locations of reconnection (e.g. Rostoker, 1996; Baker et al., 1996)? Since the onset of reconnection occurs at the end of the growth phase of substorms, it is appropriate to make an estimate of the boundary location in a stretched tail configuration with the outer part of region B quite tail-like. In appendix A, the location of the critical X line is calculated from a simplified model. The uncertainties are large because of the approximations used, but the results are consistent with a downtail distance of 15-25Re for A-B in a stretched tail situation using MHD. This distance is probably a minimum since plasma and energy losses in non-MHD situations would move the X line location even further downtail.

## 5. Reconnection at critical X lines

Consider reconnection at a near-earth neutral line. Field lines in the plasma sheet are stretched in the downtail direction and compressed in the direction perpendicular to the neutral sheet. In this situation, the boundary conditions on outflow from the earthward and tailward sides of an X line probably control reconnection (e.g. Axford, 1999). If both earthward and tailward outflow are blocked, there can be no reconnection. If outflow is blocked in only one direction, reconnection can occur if the X line moves in the other direction (i.e. there is outflow in both directions in the X line reference frame). If outflow is allowed on both sides, an X line will move so as to equalize the tailward and earthward magnetic flux transfer rates in the frame of the X line.

We now apply the above to reconnection at a critical X line in the configuration shown in figure 1b. First, consider the case where there is no flow in region B; earthward flow from a critical X line is blocked. If the plasma sheet is thick, tailward flow in region A is blocked because an outflow would stretch closed plasma-sheet field lines, resulting in a large increase in the total magnetic energy. In this case, outflow from a critical X line is blocked in both directions, and there would be no reconnection. This qualitative interpretation is probably consistent with the results of studies of the stabilizing effect of a normal component of the magnetic field on tearing-mode instabilities (e.g. Pritchett and Coroniti, 1992). The standard model of the onset of reconnection requires that the plasma-sheet become quite thin before reconnection occurs.

Now consider the case where there is an earthward flow in the outer part of region B, as shown in figure 2. The flow in region B provides an outflow on the earthward side of a critical X line. Tailward outflow occurs (in the X line reference frame) if the X line propagates toward the earth with a speed less that the earthward flow. Inside a thick plasma sheet, this results in the buildup of a magnetic island (or flux rope) on the tailward side of the X line. Figure 2a also shows the X line closing to an O line in the equatorial plane (e.g. Vasyliunas, 1998).

The convection illustrated in figure 2 is necessarily time dependent because of the growth of the magnetic island and the earthward motion of the X line. The concept of the formation and growth of a magnetic island on the tailward side of the X line is not new and is part of the standard model of substorms. The proposed earthward movement of the X line and the control of reconnection by earthward flow in region B are modifications proposed in this paper.

The remaining part of the puzzle is to understand the cause of the earthward flow in the outer part of region B. Basically it is the convection out of the nightside. Energetic plasma convects westward out of the nightside by the combined effects of curvature and gradient drift and also **ExB** drift. The plasma removal can also be described as westward motion of the eastern end of the partial ring current / plasma sheet. As a result, previously tail-like flux tubes deflate to a more dipole-like form, and, in addition, the amount of magnetic flux in the nightside of region B decreases. Both the deflation and the flux decrease result in a decrease in the radial extent of region B in the equatorial plane. The decrease occurs by earthward movement of A-B and earthward flow of flux tubes in the outer part of region B.

We now use a simplified model to determine under what conditions earthward flow in region B leads to the production of a thin current sheet in the center of the plasma sheet. Presumably, a thin current sheet leads, in turn to the onset of reconnection.

## 6. A thought experiment to illustrate the production of thin current sheets

Since simulations are beyond the scope of this paper, we assume that the production of thin current sheets leads to the onset of reconnection, and discuss the convection conditions that produce thin current sheets in the center of a plasma sheet. In particular, we look at the ability of earthward flow in the outer part of region B to produce thin current sheets.

We conduct a "thought experiment" with the simplified model shown in figure 3a. Region B and region A closed flux tubes are shown. The high-latitude boundaries (upper and lower in figure 3) are inside region A between closed and lobe field lines. We neglect the effects of transfer of magnetic flux across this boundary since reconnection at a distant neutral line is not expected to have a direct affect on near-earth reconnection. We consider an idealized situation with an earthward flow in the outer part of region B and no time dependence of the magnetic plus plasma pressure at the high-latitude boundaries; that is $P_t = P_t(x)$.

Figure 3a shows the initial configuration. Flux tube A1 cannot simply convect earthward to replace flux in region B as it flows earthward. The boundary condition that $P_t = P_t(x)$ is independent of time implies that there is little change on flux tube A1 in response to the earthward flow in region B. The plasma pressure, magnetic field intensity, and flux tube volume and length undergo little change. The flow in region A is primarily perpendicular to the neutral sheet and does not involve significant compression of the plasma. The resulting configuration is illustrated in figure 3b. The boundary A-B has moved earthward and a thin current sheet has formed. The "wings" of flux tube, A1, have moved toward the neutral sheet. Figure 4 illustrates plasma pressure profiles across the plasma sheet corresponding to figures 3a and 3b. The pressure profile across the two wings of region A does not vary significantly as they convect toward the center of the plasma sheet since their total magnetic plus plasma pressure is already equal to the pressure at the lobe boundary. However, the plasma pressure at the central plane must be approximately equal to $P_t = P_t(x)$. This requires the formation of a thin intense current sheet in the center of the plasma sheet in the absence of reconnection, as shown in 3b. Figure 3c illustrates the expected configuration if reconnection occurs.

The author believes that the special case in the thought experiment simulates the situation when the earthward flow in region B is large and the rate of transport of magnetic flux into the tail is small. Thin current sheet formation and reconnection are favored. This condition is consistent with a well-developed growth phase and a northward turning of the interplanetary magnetic field.

The importance of a northward turning is not clear from the above. The author has not been able to illustrate the effect of strong or weak dayside reconnection with a simple model. The studies presented in Erickson (1984) may be relevant. In the cases in that study, the lobe pressure is increasing, implying that dayside reconnection and flux transport into the tail is strong. The thick plasma sheets become thinner as the tail pressure increases, but there is no indication of the formation of a thin current sheet in the center of the plasma sheet. Further simulation studies are needed to see if increasing magnetotail pressure "suppresses" the formation of thin current sheets inside the plasma sheet. A suppression effect would provide a simple explanation of substorm triggering by northward turnings of the interplanetary magnetic field, however it is not an essential part of the model presented here.

In summary, the thought experiment has indicated that earthward flow in the outer part of region B tends to produce a thin current sheet in the center of the plasma sheet, and hence favors the onset of reconnection. The earthward flow is the result of the combined effects of dipolarization of region B flux as the eastern end of the partial ring current drifts past a given local time, and of the decrease in the total amount of region B magnetic flux on the nightside due to convection. It follows that reconnection is associated with the eastern end of the plasma sheet / partial ring current.

## 7. Evolution of the reconnection and Bursty Bulk Flows

After onset at a critical X line at a given local time, the reconnection evolves into lobe merging. The simplest model for this evolution is that it eats through the plasma sheet field lines until the lobe is involved.

According to the description in section 5 of this paper, the X line propagates earthward initially. If the earthward flow of the outer part of region B involves an earthward movement of the boundary, A-B, ~10Re, the X line would be expected to move approximately half this distance. If the reconnection produces a component normal to the equatorial plane ~10nT, the amount of flux reconnected would be approximately 0.3 webers per meter of cross-tail distance. This is approximately the amount of flux in a 3Re thick plasma sheet, and it can be concluded that this initial phase reconnects a substantial fraction of the plasma sheet.

If we assume that that auroral breakup is related to the reconnection of flux at the X line, the above phase of the reconnection takes a few minutes at a given local time. From this, the reconnection rate is ~.001 volts/meter and the flow velocity in the X line reference frame is ~200 km/sec (400 km/sec in the stationary frame). This is consistent with the observed bursty bulk flow velocities, as well as with the magnetic flux transfer rate (e.g. Baumjohann et al., 1990; Schodel et al., 2001). Finally, a width of several earth radii is obtained for the reconnection region if we multiply the drift velocity of the east end of the partial ring current (several tens of kilometers per second) by the duration of the event at a given local time (a few minutes). This width is also consistent with bursty bulk flows. More generally, the model appears to be consistent with the conclusions of Angelopoulous et al., 1999, that the flows are responsible for much of the transport of magnetic flux to the nightside from the tail.

It follows that the eastern end of the drifting plasma (partial ring current) has a dimension of a few earth radii or possibly less. The author is not aware of any published values for the scale size of the eastern end of the partial ring current. It seems quite likely that there is a steepening process at work; the slower drifting particles are caught up in the earthward flow of flux tubes and are energized, giving them higher drift rates. This needs further study.

After onset, the reconnection slot would be expected to propagate westward with the eastern end of the drifting energetic plasma. This westward motion provides a ready explanation of the westward traveling surge (Akasofu, 1964).

## 8. Summary and Discussion

The convection of energetic ions (curvature and gradient drifts as well as **ExB** drifts) and of magnetic flux out of the nightside is an important boundary condition for reconnection at near-earth neutral lines. Hence it is necessary to study nightside near-earth reconnection in the context of a convecting magnetosphere. The simplified model of the magnetosphere shown in figure 1 is used in this study. It consists of three regions of magnetic flux: 1) region A consisting of tail lobe plus tail-like plasma sheet flux which cannot convect to the dayside without loss of energy (by reconnection), 2) region B consisting of magnetic flux which contains energetic plasma and which can convect to the dayside, 3) region C consisting of lower-latitude flux tubes without significant plasma energy.

Magnetic flux is transferred out of region A into region B by nightside reconnection at a "critical" X line located at the boundary between A and B. We defined the regions such that reconnection must occur at a critical X line (either continuously or spasmodically) for convection from the tail to the dayside to be possible. An estimate of the location of the critical X line gives a value of 15 to 25 Re downtail; however, since it is based on MHD, the estimate is believed to be a minimum and larger values are likely.

Reconnection at a critical X line cannot occur if outflow from the X line is blocked in both downtail and earthward directions. However, if there is an outflow toward the earth in the outer part of region B, reconnection can occur (see figure 2). Propagation of the X line toward the earth allows both earthward and tailward flow in the reference frame of the X line. A magnetic island grows on the tailward side of the X line.

The earthward flow in the outer part of region B is the result of a combination of deflation of tail-like flux tubes in region B as energetic plasma drifts westward out of the nightside, and **ExB** drift out of the nightside. Both effects result in a thinning of region B (a decrease in the down-tail extent) and produce an earthward flow in the outer part of region B. The removal of plasma energy from a given local time can be described as the westward motion of the eastern end of the partial ring current and plasma sheet.

A thought experiment was developed to show that an earthward flow in region B tends to produce a thin current sheet in the center of the plasma sheet, which favors the onset of reconnection. Hence a strong convection to the dayside, as occurs in the late growth phase of substorms, favors the production of a thin current sheet and the onset of reconnection in the center of the plasma sheet. The effect of dayside reconnection is not clear, although it is possible that strong dayside reconnection suppresses the formation of a thin current sheet in the center of the plasma sheet. If true, this would explain triggering by northward turnings of the IMF.

The explanation of substorm expansions is that earthward flow occurs in the outer part of region B after the curvature and gradient and **ExB** drift velocities have increased in the growth phase. The earthward flow and onset of reconnection occur in a relatively narrow local-time slot at the eastern end of the partial ring current / plasma sheet. The reconnection evolves into lobe merging with a time scale of several minutes at a given local time. The X line first moves earthward and then tailward as it evolves toward lobe merging. The reconnection slot moves westward across the tail with the east end of the drifting plasma, accounting for the westward traveling surge.

The above model is topologically the same as the standard model, but differs in the specific mechanism it provides for the onset of near-earth reconnection, the behavior of the X line, and the westward propagation of the reconnection slot.

The limited east-west extent of substorm features such as bursty bulk flows, the initial auroral breakup (Akasofu, 1964), particle injection at geosynchronous orbit (Arnoldy and Moore, 1983) and the dipolarization (Nagai, 1982)), can be explained as the scale size of the eastern end of the plasma distribution. It was noted that there is probably a steepening effect of the east end because the slower-drifting particles are energized by the earthward flow. Other observations explained include the behavior of X lines. Since the X lines first travel earthward and then tailward in the lobe-merging phase, both tailward and earthward motions would be expected, as reported by Ueno et al. (1999).

There are several earlier works that considered control of substorm onset by changes in the configuration of the coupled dipole-like magnetosphere and near-earth plasma sheet. Petschek and Siscoe (1996) suggested that the breakup was related to the convection rate to the dayside. Atkinson (1991) proposed that a major northward turning of the IMF near the end of the growth phase required a reconfiguration of the convection and of the Alfven layers, and hence of magnetosphere topology, leading to interchange instability with reconnection. Lyons (1995) proposed a model in which a northward turning produces changes in the convection pattern which results in changes in the plasma pressure distribution in the near-earth plasma sheet causing substorm expansion.

Another class of theories that is related to the work here involves east-west displacement of plasma energy on high-beta flux tubes. These include interchange and ballooning instabilities (e.g. Roux et al., 1991; Cheng and Lui, 1998). Most of these studies look for an instability of an initially uniform distribution of plasma across the nightside, but it is possible that an instability of this type plays a role at the eastern end of the partial ring current plasma. The suggested steepening and westward drift of the east end of the partial ring current are behavior which occurs in ballooning theories.

## Appendix A: Estimate of distance to the critical X line and boundary A-B.

To be able to convect to the dayside, a flux tube must arrive on the dayside with $\beta_{eq} \leq 1$ ($\beta_{eq}$ is the ratio of plasma to magnetic energy density in the equatorial plane). If this condition is not satisfied, the flux tube is too tail-like and remains trapped in the tail as part of the region A flux. Thus the boundary A-B is defined by flux tubes which can arrive at the dayside with $\beta_{eq} = 1$ (the value on the nightside may be greater). We denote nightside plasma-sheet values of variables by the subscript, PS, and dayside values by the subscript, D. The condition that $\beta_{eq} = 1$ on the dayside magnetopause is given by $P_D = B_D^2 / 2\mu_0$. We use average values of the variables $P_{PS}$ and $B_{PS}$ along a field line to define a value, $\beta_{PS}$, such that $P_{PS} = \beta_{PS} B_{PS}^2 / 2\mu_0$. Using these relationships and assuming adiabatic convection with γ=5/3, the ratio of the volume of a boundary flux tube in the nightside plasma sheet to the volume on the dayside, is given by

$$\frac{V_{PS}}{V_D} = \left(\frac{P_D}{P_{PS}}\right)^{\frac{3}{5}} = \left(\frac{B_D}{\sqrt{\beta_{PS}}\,B_{PS}}\right)^{\frac{6}{5}} \qquad (1)$$

Generally $B_D / B_{PS} > 1$. For values $1 < B_D / B_{PS} < 4$ and $\beta_{PS} = 1$ we find the range of values $1 < V_{PS} / V_D < 5$. For $\beta_{PS} = 2$, the range is $0.7 < V_{PS} / V_D < 3.3$. It should be remembered that the value of $\beta_{PS}$ is an average value along a field line and not the equatorial value, so that the above values are probably reasonable.

We can use equation 1 to obtain a very approximate value for the distance to the critical X line and boundary using assumptions appropriate for a stretched tail. We assume that the volumes of flux tubes on both the nightside and the dayside can be approximated by the volume of their outer parts (that is we neglect the volume of the dipole-like part of the flux tubes). On the dayside, the volume is given by $V_D \approx L_D / B_D$, where $L_D$ is length of the magnetopause between cusps, and $1 / B_D$ the cross-sectional area of the tube. On the nightside, the volume is given by $V_{PS} \approx L_{PS} / B_{PS}$, where $L_{PS}$ is twice the distance from the outer limit of dipole-like field lines (at ~10 Re) to the boundary A-B and critical X line site. Substituting the above values in equation 1 gives

$$\frac{L_{PS}}{L_D} = \left(\beta_{PS}\right)^{-\frac{3}{5}}\left(\frac{B_D}{B_{PS}}\right)^{\frac{1}{5}} \tag{2}$$

For $1 < B_D / B_{PS} < 4$, $\beta_{PS} = 1$ and $L_D = 20\,\text{Re}$, we find a range $20 < L_{PS} < 26\,\text{Re}$. For $\beta_{PS} = 2$, the range of values is 14 to 17 Re. The above values suggest that the boundary and X line are at a distance from the earth between 15 and 25 Re for MHD convection. As stated earlier, this estimate assumes a stretched tail-like topology and it is probably most valid near the end of a substorm growth phase. If particles are lost from flux tubes during the convection, the X line would be further downtail.

**Acknowledgment**—This research was supported by a grant from NSERC.

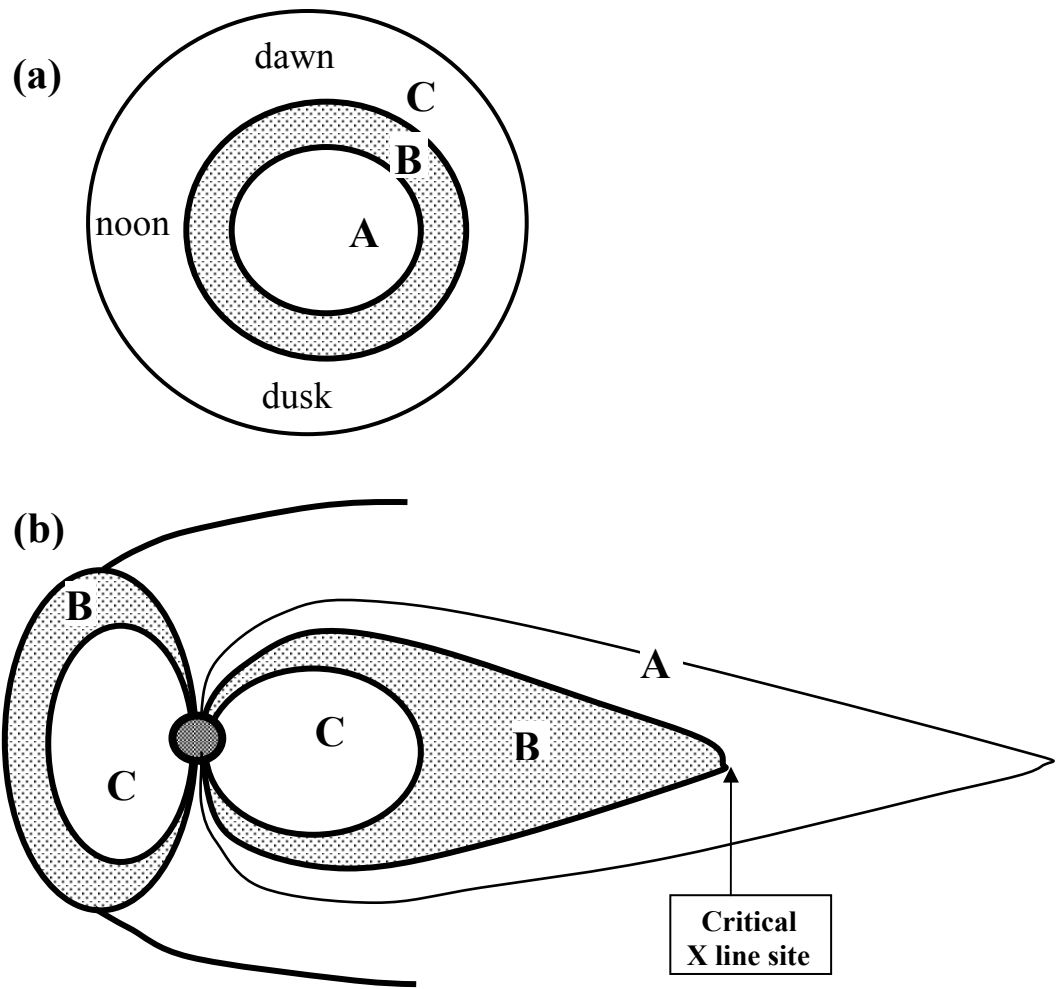

Figure 1. The three-region magnetosphere: (a) polar view at ionospheric heights, (b) noon-midnight section. Region A—tail-lobe plus plasma-sheet magnetic flux that cannot convect to the dayside; Region B—plasma sheet/ring current flux that can convect to the dayside; Region C—magnetic flux with low plasma energy density

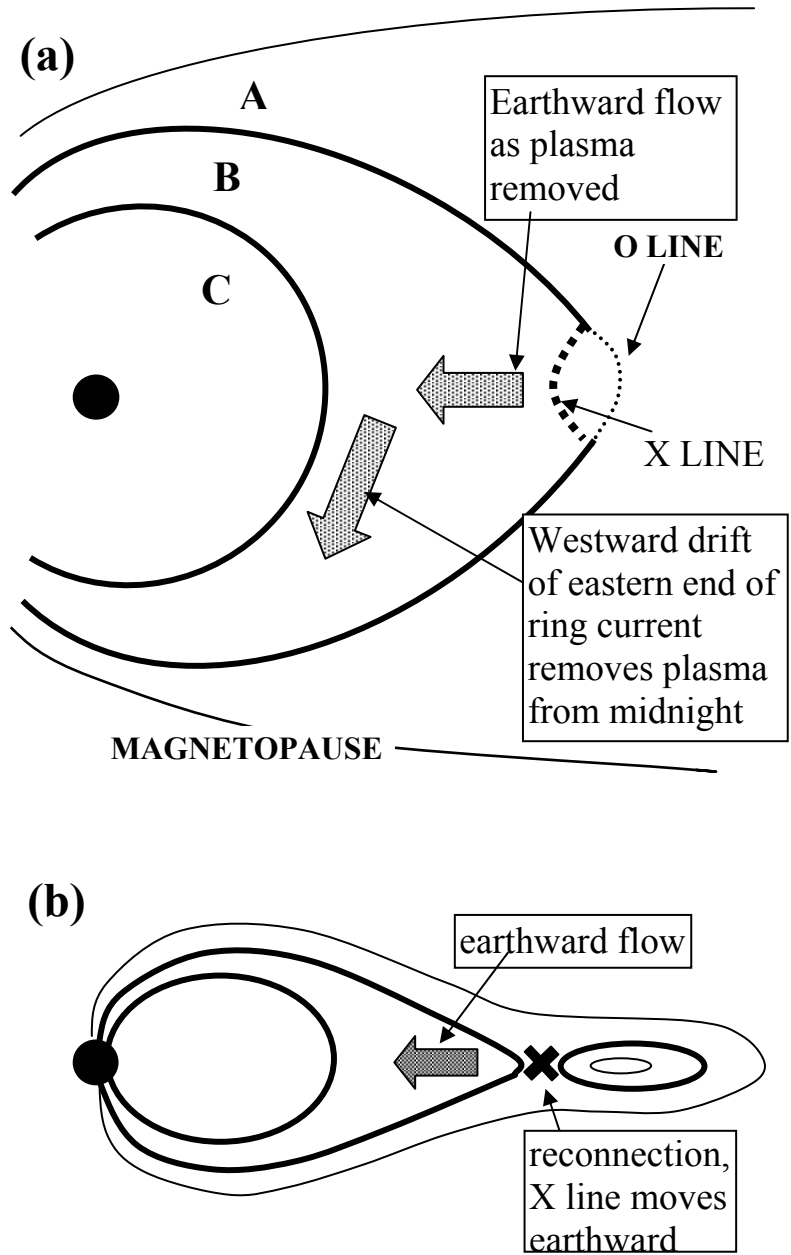

**(a)**

A

B

C

Earthward flow as plasma removed

**O LINE**

**X LINE**

Westward drift of eastern end of ring current removes plasma from midnight

**MAGNETOPAUSE**

**(b)**

earthward flow

reconnection, X line moves earthward

Fig 2.  Reconnection at a critical X line inside a thick near-earth plasma sheet: (a) equatorial plane (b) midnight section.  The combined curvature and gradient plus **ExB** drifts remove energetic plasma and magnetic flux from the nightside, and thus reduce the equatorial thickness of region B (deflation of tail-like tubes plus an equipotential contribution).  The resulting earthward flow of the outer part of region B provides outflow on the earthward side of an X line at the boundary A-B.  Reconnection can then occur if the X line moves earthward and an island forms tailward of the X line.

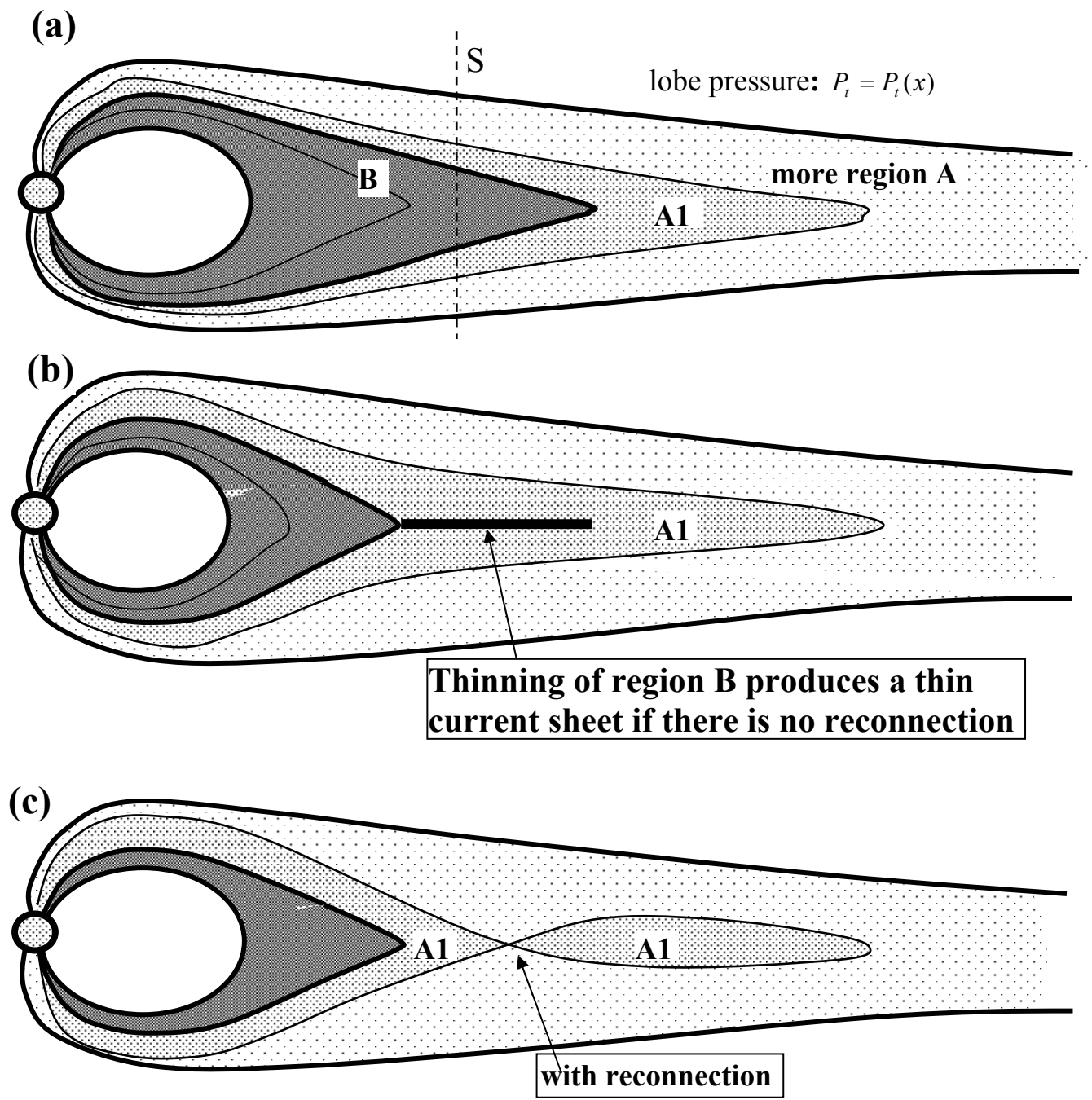

**(a)**

lobe pressure: $P_t = P_t(x)$

S

B

more region A

A1

**(b)**

A1

**Thinning of region B produces a thin current sheet if there is no reconnection**

**(c)**

A1

A1

**with reconnection**

Fig 3. Midnight section showing: (a) the initial configuration, (b) the configuration after the earthward flow of flux in region B, and (c) the configuration if flux tube A1 reconnects.

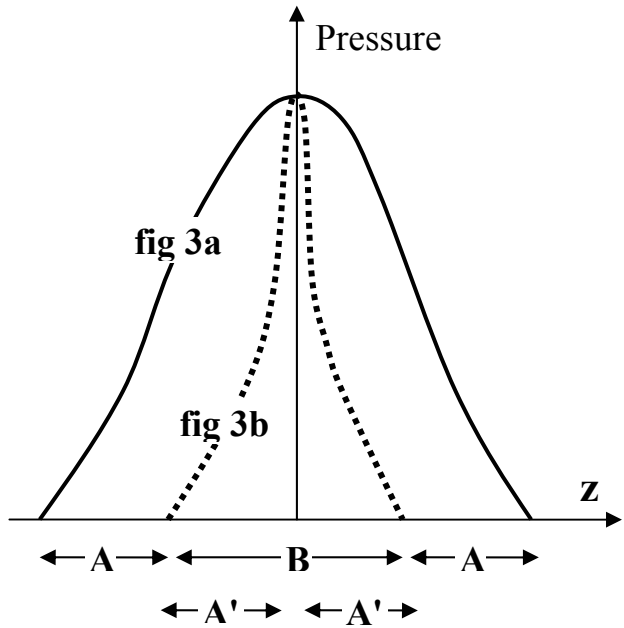

Figure 4. Pressure versus distance from the neutral sheet at tail cross section S. The solid line corresponds to figure 3a and the dotted line to figure 3b. The north-south extents of the regions are indicated by A and B (for figure 3a) and by A' (for figure 3b).